\documentclass{jpp}
\usepackage{graphicx,wrapfig}
\usepackage{epstopdf, epsfig}
\usepackage{natbib}
\usepackage{hyperref}
\usepackage{caption}
\usepackage{subcaption} 

\usepackage{color}
\usepackage{amsmath}

\shorttitle{WiPAL}
\shortauthor{Cary B. Forest et al.}

\title{The Wisconsin Plasma Astrophysics Laboratory}

\author{C.B. Forest\aff{1} 
\corresp{\email{cbforest@wisc.edu}}, 
K. Flanagan\aff{1},
M. Brookhart\aff{1},
M. Clark\aff{1}, 
C.M. Cooper\aff{1},
V. D\'{e}sangles\aff{2}, 
J. Egedal\aff{1}, 
D. Endrizzi\aff{1}, 
I. V. Khalzov\aff{3},
H. Li\aff{4},
M. Miesch\aff{5}, 
J. Milhone\aff{1},
M. Nornberg\aff{1}, 
J. Olson\aff{1},
E. Peterson\aff{1},
F. Roesler\aff{1}, 
A. Schekochihin\aff{6,7},
O. Schmitz\aff{1},
R. Siller\aff{1},
A. Spitkovsky\aff{8},
A. Stemo\aff{1},
J. Wallace\aff{1}, 
D. Weisberg\aff{1} 
\and E. Zweibel\aff{1} }

\affiliation{
\aff{1} Department of Physics,  University of Wisconsin-Madison,
Madison, WI 53706 USA 
\aff{2} Laboratoire de Physique de l'Ecole Normale Sup\'erieure de Lyon, CNRS and Universit\'e de Lyon, 46 all\'ee d'Italie, 69364 Lyon Cedex 7, France
\aff{3} National Research Centre ``Kurchatov Institute", Moscow, 123182, Russia
\aff{4} Los Alamos National Laboratory, Los Alamos, NM 87545, USA
\aff{5} High Altitude Observatory, National Center for Atmospheric Research, Boulder, CO 80307-3000, USA
\aff{6} Rudolph Peierls Centre for Theoretical Physics, University of Oxford, Oxford OX16NP, UK 
\aff{7} Merton College, Oxford OS14JD, UK
\aff{8} Department of Astrophysical Sciences, Princeton University, Princeton, NJ 08540, USA
}

\begin{document}

\maketitle

\begin{abstract}
The Wisconsin Plasma Astrophysics Laboratory (WiPAL) is a flexible user facility designed to study a 
range of astrophysically relevant plasma processes as well as novel geometries that mimic astrophysical systems. A multi-cusp magnetic bucket constructed from strong samarium cobalt permanent magnets now confines a 10 m$^3$, fully ionized, magnetic-field-free plasma in a spherical geometry. Plasma parameters of $ T_{e}\approx5$ to $20$ eV and $n_{e}\approx10^{11}$ to $5\times10^{12}$ cm$^{-3}$ provide an ideal testbed for a range of astrophysical experiments including self-exciting dynamos, collisionless magnetic reconnection, jet stability, stellar winds, and more. This article describes the capabilities of WiPAL, along with several experiments, in both operating and planning stages, that illustrate the range of possibilities for future users. 
\end{abstract}

\title[WiPAL]{The Wisconsin Plasma Astrophysics Laboratory}

\section{Introduction}
\begin{figure}
  \centerline{\includegraphics[scale=1]{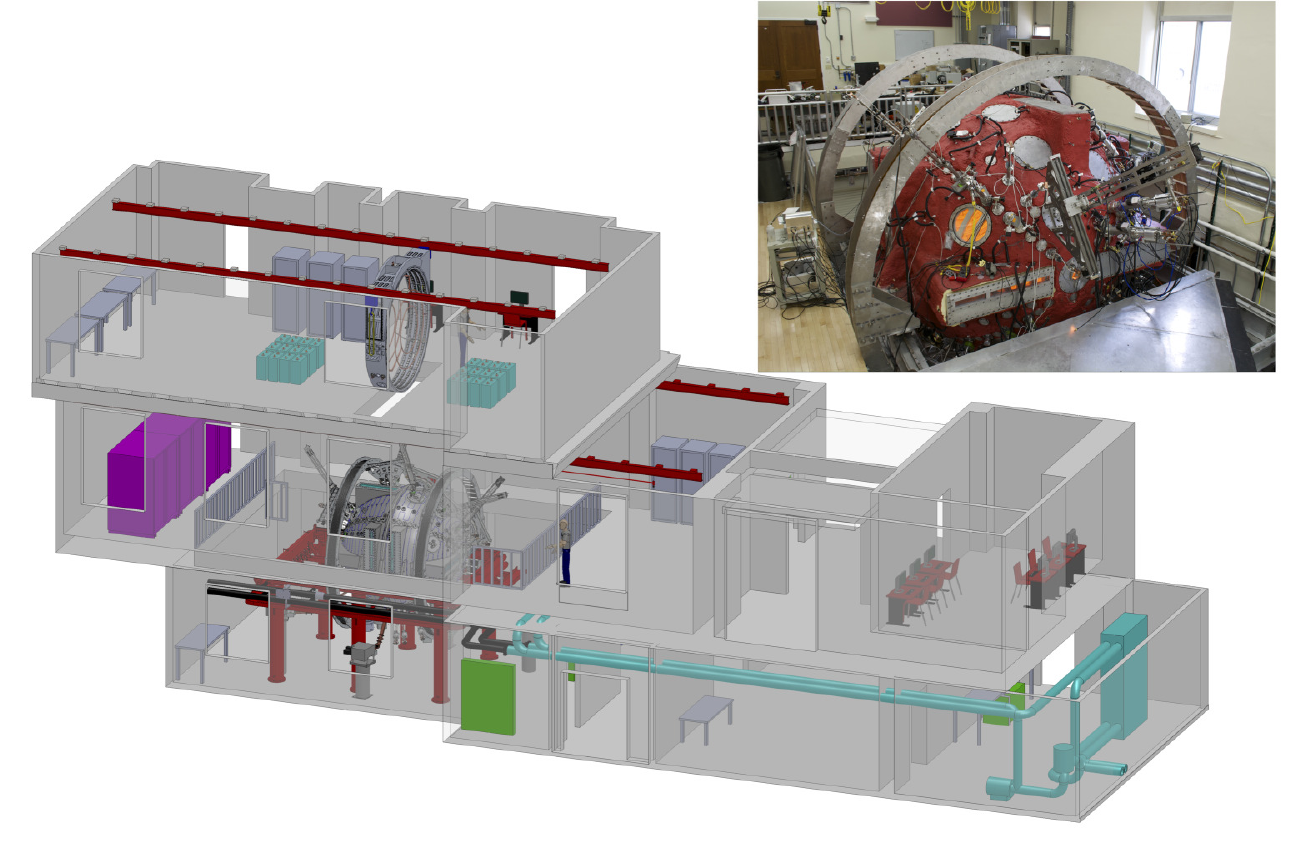}}
  \caption{The WiPAL facility (left) consists of a main laboratory space where the vessel is housed and several auxiliary spaces for high voltage power management, water pumping, computer control, and housing the TREX insert. Photo in upper right shows the 3 m diameter vessel covered with arrays of probes and clear viewports.}
  \label{fig:WiPAL1}
\end{figure}

During the past five years, a medium-scale multi-investigator experimental plasma facility---the Wisconsin Plasma Astrophysics Laboratory (WiPAL)---has been constructed and is now in operation at the University of Wisconsin-Madison.
At the heart of the facility is a high bay that houses a 3 m diameter spherical multi-cusp confinement device (see figure~\ref{fig:WiPAL1}).
Its flexibility and diagnostic access allow WiPAL to currently support two major experiments: the Madison Plasma Dynamo Experiment (MPDX) and the Terrestrial Reconnection Experiment (TREX), along with several new experiments being added in the near future. Both the flexible design of the facility and the shared hardware between experiments allow for a rapid turn-around between different configurations.

\begin{wrapfigure}{r}{2.5in}
  \includegraphics[scale=1]{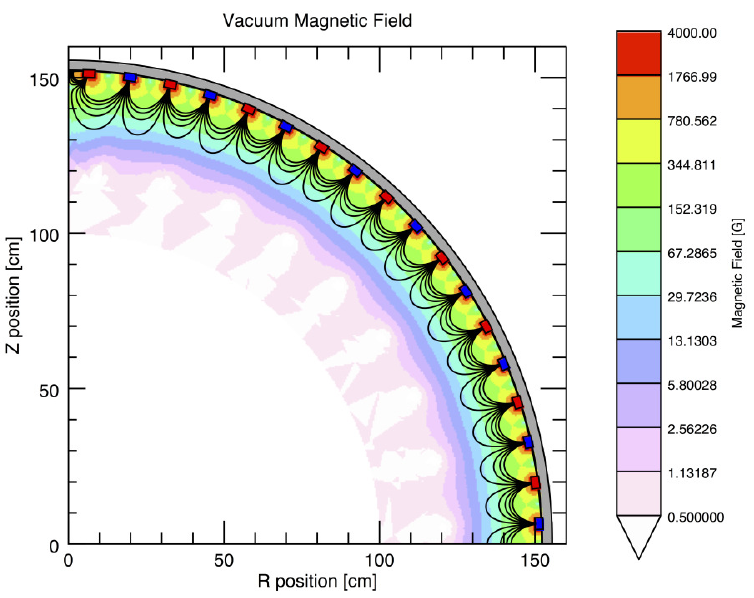}
  \caption{A static calculation of the vacuum magnetic field from the
    permanent magnets at the edge and the background earth field. This
    calculation has been normalized by hall probe measurements along a
    radial chord.}
      \vspace{-.1in}
  \label{fig:Cusp}
\end{wrapfigure}

The purpose of this paper is to describe the physical parameter regimes that can be achieved in WiPAL, to illustrate its flexibility, and to show how the plasma confinement scheme lends itself to many experiments of interest to the astrophysics and fundamental plasma physics communities. Already, this facility has transitioned from a single-purpose experiment focused on plasma dynamos (MPDX) to a multi-investigator facility accommodating
a new reconnection experiment (TREX). In the future,  WiPAL will transition into a collaborative user facility open to other laboratory astrophysics research groups. This paper illustrates the potential of the facility by describing (in addition to MPDX and TREX) several additional experiments 
that are now being pursued:  acoustic and Alfv\'{e}n wave propagation in connection with helioseismology; pulsar and stellar wind launching from a rotating dipolar magnetosphere; jet formation and propagation into background plasma; and small-scale high-power helicity injection. In addition to these experimental efforts, numerical and theoretical computations, often carried out with identical dimensionless parameters, are used to create predictive models and inform elements of design.

This article begins with an overview of WiPAL along with experimentally achieved plasma parameters. Then a description is given of the diagnostic suite and of the magnetic cusp confinement used in WiPAL.  This is followed by brief summaries of experiments that are under way or in development.

\section{Description of the Facility}

The WiPAL facility (shown in figure~\ref{fig:WiPAL1}) consists of the plasma confinement vessel and the associated infrastructure (high bay, electrical power, water cooling, and associated laboratories). The vacuum vessel consists of two 1.5 m radius cast-aluminum hemispheres mounted on a track that can be separated during a vacuum opening. This provides the opportunity for quickly inserting different devices inside the WiPAL vessel.  For example, a transition from full MPDX operation to the TREX configuration requires only a few days.

Nearly 3000 samarium cobalt (SmCo) permanent magnets (each with $|{\bf B}|>3$ kG at their surface) are held in axisymmetric rings on the inside of the vessel. These rings alternate in polarity to form a strong high-order multipole field that decreases in strength to the background earth field within 20 cm of the vessel wall (shown in figure~\ref{fig:Cusp}). This edge-localized cusp field provides sufficient plasma confinement to achieve the parameters listed in table~\ref{tab:param} while leaving the ions in the core unmagnetized. The rings were carefully produced 
to assure axisymmetry, thus reducing many aspects of WiPAL experiments to two dimensions. Since all the magnetic field variation is in the radial and polar directions, gradient and curvature drifts are azimuthal. This leads to an azimuthal symmetry of the generated plasmas, thereby reducing convective losses associated with magnetic ripple. 

\begin{figure}
  \centerline{
  \includegraphics[scale=1]{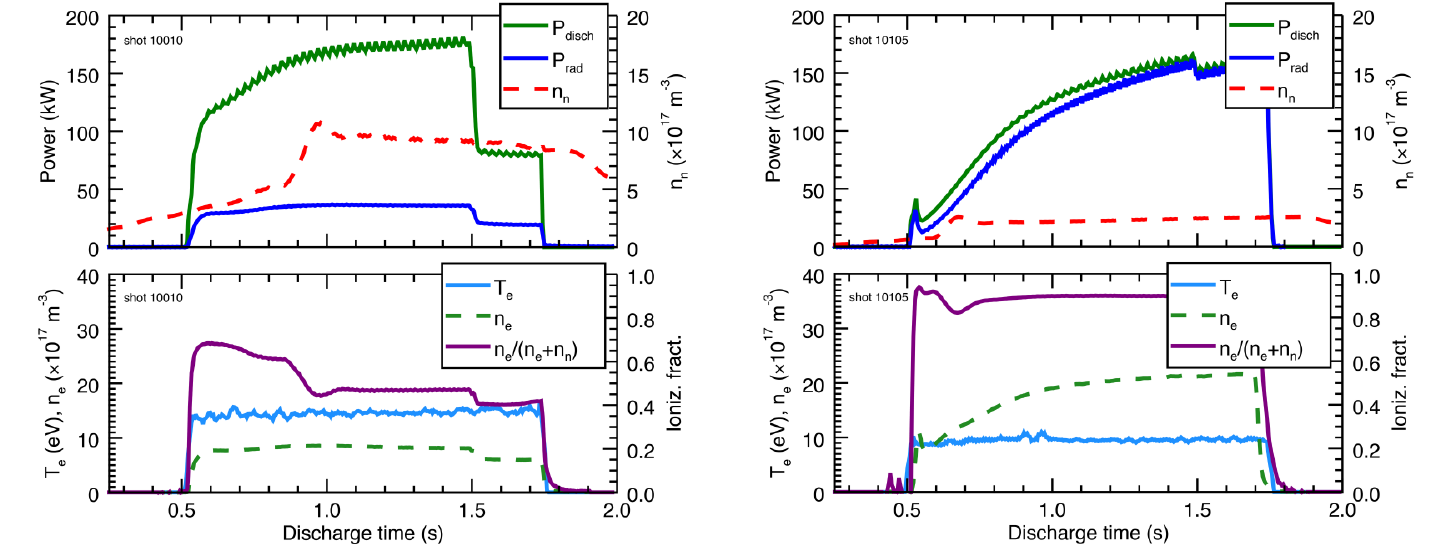}}
  \caption{Time traces of plasma parameters and input power from a typical WiPAL discharge in helium (left) and argon (right). Radiated power is measured with a bolometer, $n_{n}$ is measured with a pressure gauge at the edge of the vessel (assuming room temperature neutrals), $T_{e}$ is measured with a Langmuir probe inserted into the unmagnetized bulk, and $n_{e}$ is measured with the mm-wave interferometer system.}
  \label{fig:Plasma1}
\end{figure}

\begin{wraptable}{r}{2.925in}
\def~{\hphantom{0}}
\begin{tabular}{l || c | c | c |}
Parameter&Achieved&Achieved&Projected\\
Gas&He&Ar&He \\
\hline
\hline
$P_{in}(kW)$&300&300&650\\
$T_{e}(eV)$&20&10&40\\
$T_{i}(eV)$&1.5&3&10\\
$n_{e}(10^{12}$ cm$^{-3})$&2.5&4&10\\
$f_{\%}$&75&90&99\\
$V$ (km s$^{-1}$)&10&5&20\\
$Rm=VL/\eta$&900&350&5000\\
$Re=VL/\nu$&600&1300&1000 \\
$B_{equip}$ (G)  & 14  & 29 & 60 \\
$\beta$ at 10 g & 8 & 8 & 20 \\
\end{tabular}

\caption{Table of parameters achieved to date in separate discharges (with 10 of 18 possible cathodes and no ECR heating) as well as projected parameters. The ionization fraction, $f_{\%}$, is calculated using neutral pressure at the vessel wall and line-integrated electron density.}
\label{tab:param}
\end{wraptable}
To date, argon and helium plasmas have been created and heated by an array of hot emissive lanthanum hexaboride (LaB$_6$) cathodes, each independently powered by separate 30 kW power supplies (currently 12, with 18 planned). See figure~\ref{fig:Plasma1} for an example discharge in a typical helium plasma. Five 20 kW magnetrons at 2.45 GHz are in the process of being added to independently heat electrons through electron cyclotron resonance (ECR) heating in the magnetized edge, bringing the maximum steady-state input power to $\sim650$ kW. These sources are quasi-stationary, meaning that plasmas can be sustained for tens of seconds limited only by vessel heating.

A unique capability of the facility is the ability to control the plasma rotation at the boundary \citep{MPDXPOP}.  Early experiments at the facility have been focused on studying highly conducting, flow-dominated plasmas and understanding the interface between magnetized and unmagnetized plasma conditions. Plasmas have been stirred at speeds up to 10 km s$^{-1}$ in the azimuthal direction by applying a ${\bf J}\times{\bf B}$ torque in the magnetized edge region. Each of the cathodes are mounted on a motorized stage so their insertion depth can be quickly and accurately set between shots. Cathodes pushed into the unmagnetized bulk of the plasma ionize and heat the plasma, while cathodes pulled back into the magnetized region draw current across magnetic field lines, injecting torque into the plasma. This torque is viscously coupled to the unmagnetized core. This boundary-driven flow scheme was designed for dynamo studies on MPDX, but will be used to study other features of fast-moving plasmas in the WiPAL facility.

Additionally, several capacitor banks with ignitron switching have been used to drive short pulse reconnection experiments in TREX (\S\ref{sec:TREX}) and to inject current for plasma jet experiments (\S\ref{sec:jets}). For example, initial reconnection experiments utilized an up to 10 kV, 100 $\mu$F capacitor bank to drive approximately 20 kA for 250 $\mu$s, generating a 10 G reconnection field from a two-turn internal coil. WiPAL is also equipped with a large Helmholtz coil pair which provides a uniform field throughout the plasma volume of up to 275 G. 

\subsection{Major Diagnostics}

The WiPAL vessel is covered with approximately 200 ports for diagnostic access of plasma discharges via robotic probe arrays and advanced optical measurements. Owing to the high temperature and steady-state nature of WiPAL discharges, {\it in situ} probes have been carefully designed to accommodate high heat flux. Additionally, the small loss area provided by the cusp confinement is increased significantly by inserting probes into the plasma. Motivated by these issues, a suite of non-invasive optical diagnostics has been developed and works in concert with inserted probes. 

\subsubsection{Two-dimensional Probe Systems}\label{sec:Probes}
\begin{wrapfigure}{r}{2.5in}
\vspace{-.2in}
  \centerline{\includegraphics[scale=1]{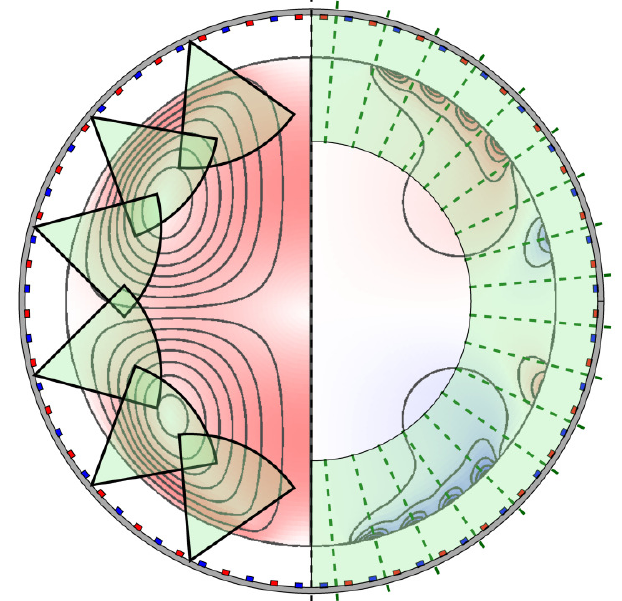}}
  \caption{An example of a predicted dynamo-capable axisymmetric flow, with poloidal flow (left) and toroidal flow (right) is well diagnosed by the full motorized probe suite. WiPAL probe coverage will include both sweep probes (6 total, left) and linear probes (18 total, right).}
  \vspace{-.2in}
  \label{fig:probecoverage}
\end{wrapfigure}

Most plasma astrophysics experiments require high-resolution, two-dimensional (2D) measurements obtained using internal probe arrays.  
To meet this need, WiPAL has several robotically controlled probe drive systems in place. Two types of scanning systems are used, as shown in figure
~\ref{fig:probecoverage}.
The first probe system is a set of sweeping probes (one currently installed, with six planned) that are inserted through a single ball-joint vacuum seal port, combining radial and angular motion to sweep out a 2D plane. Two stepper-motor-controlled transverse stages allow the regions shown to be scanned shot-to-shot with arbitrary resolution.  The second probe system is an array of single-axis probes that are inserted every 10$^{\circ}$ in latitude. These probes have a single stepper-motor drive, allowing radial shot-to-shot scans.

The main probe type to be used in WiPAL is a combination velocity-magnetic probe. Each probe tip consists of two Mach probes (four planar Langmuir faces collecting ion saturation current) to measure two orthogonal components of $ \mathbf{v}$. Development is under way to add three Hall sensors to measure the entire three-dimensional (3D) $\mathbf{B}$ field in each probe tip. These probes are designed to withstand long, warm plasma pulses, utilizing a thermally conductive copper shaft paired with an electrically and thermally insulating quartz shield. The lifetime of probes is ultimately limited by accumulated heat load over many plasma pulses, but thermocouple temperature monitoring facilitates optimization of plasma duty cycle versus probe temperature. 

Increased probe coverage also leads to diminished plasma performance because plasma losses to the exposed probe area can be much larger than the edge losses through the permanent magnet cusp field. Motorized probes allow for flexibility in controlling the plasma loss area. For example, staggered radial scans of various probes can keep the loss area constant while still mapping out the full extent of probe coverage. 

The 2D structure of the magnetic field on the surface of the plasma will be measured using an array of 64 external magnetic probes placed in the region between the plasma and the vessel wall. These probes will consist of three-axis Hall probes for measuring low-frequency magnetic fields as well as two orthogonal {\sl Mirnov} coils for measuring higher-frequency fluctuations. The final implementation of this full array will resolve $\ell$=8 harmonics in the polar direction and $m$=8 harmonics in the toroidal direction.

In addition to this collection of internal probes, WiPAL is outfitted with an array of optical diagnostics. A millimeter-wave interferometer system measures the line-integrated electron density; a Fabry-Perot spectrometer provides measurements of ion temperature; and optical emission spectroscopy (OES) can be used to determine neutral and ion density profiles, electron temperature, and electron density. All of these systems provide precise, non-invasive measurements of key plasma parameters without increasing the plasma loss area.

\subsubsection{Absolute Density Measurements} WiPAL is equipped with a heterodyne millimeter-wave (320 GHz) interferometer system designed to measure chord-integrated absolute electron density via wave phase shifts. This system is very similar to interferometers used on the Helically Symmetric Experiment (HSX) and the Madison Symmetric Torus (MST) \citep{Deng2003, Deng2006}.Time resolution of density measurements is set by the intermediate frequency (IF) of the variable-frequency source. Typically, the IF is set to $f\sim1$ MHz, but is adjustable from 0.1-100 MHz. 

At WiPAL densities of $n_{e}=10^{11}$ to $5\times10^{12}$ cm$^{-3}$, phase
shifts of several fringes (one fringe corresponds to a $2\pi$ phase
shift between the reference and plasma beams) must be measured. Additionally, owing to time resolution requirements, this measurement must be made at frequencies up to several megahertz. This
constraint is met by using a high-speed field-programmable gate array (FPGA) programmed to compute the
phase at each period of the reference beam. The ultimate density
resolution is set by the speed of the FPGA clock relative to the IF
frequency and yields $\delta n\sim10^{10}$ cm$^{-3}$ for an IF of 1 MHz. 

This absolute electron density measurement is used to calibrate Langmuir probe data for point measurements as well as to constrain neutral emission models for spectroscopic measurements of the electron temperature (\S\ref{sec:OES}). Because of the good time and density resolution, this diagnostic will be used in TREX to measure density fluctuations associated with magnetic reconnection. 

\subsubsection{Ion Temperature and Flow}\label{sec:FP}
Doppler-shifted line emission is measured using a Fabry-Perot spectrometer. Plasma light is passively collected from either the $\lambda=488$nm argon ion line or the $\lambda=468.6$nm helium ion line complex depending on the working gas. By imaging these ion lines at a high resolution, the Fabry-Perot captures the line-integrated ion velocity distribution function. Ion temperature is inferred from the thermal broadening of the lines, while velocity can be measured via the Doppler shift of the peak away from its stationary value. 

The Fabry-Perot system has several distinct advantages over a grating-type spectrometer. For a comparable sized grating, the Fabry-Perot has an increase of nearly 100 times the resolving power. Additionally, because the output of the Fabry-Perot is a symmetric ring structure where $r^{2}\propto\lambda$, integration around the rings can greatly increase the signal-to-noise ratio. This procedure, called ring summing, results in a sensitivity gain of 10-30 compared to linear cuts of the pattern \citep{Coakley1996}. Additionally, it allows for relatively short integration times (roughly 1 s in this system) using a standard digital camera. WiPAL also has a high-performance intensified charge-coupled device (CCD) camera capable of taking even shorter exposures (approximately 0.1s) for multiple measurements during discharges. The typical resolution (i.e., error bars) of Fabry-Perot ion temperature measurements is $\delta T_{i}\approx0.05$ eV and the velocity resolution is of the order $\delta v\approx10$ m s$^{-1}$. 

\subsubsection{Electron Temperature and Distribution}\label{sec:OES}
Several survey spectrometers are used to measure emission throughout the visible spectrum with lower spectral resolution than the Fabry-Perot. The spectrometers capture segments of the near-ultraviolet to the near-infrared range, either monitoring the entire spectrum at low resolution (300 nm $-$ 888 nm at 0.387 nm) or small segments at a higher resolution (381 nm $-$ 511 nm at 0.045 nm). Spectra are continuously captured through WiPAL discharges with exposure times of 2-80 ms and rates of 10-50 Hz which provide a time history relevant for long-pulse time scales.

The OES system, calibrated by the millimeter-wave absolute density measurement, can estimate the electron temperature and ionization fraction. Line ratios are compared to collisional radiative models calibrated to data from other experiments and used to infer  $T_e$ while including additional corrections due to non-Maxwellian electron distribution functions in argon plasmas \citep{Boffard2010, Wang2013}. In helium, line ratios can also be used to determine $T_e$ and $n_e$ with similar modeling \citep{Schmitz2008}. Good agreement has been found between the models and the data in the regimes calibrated for tokamak scrape-off layers (high $T_e$, $n_e$). Work is under way to benchmark these models for more relevant parameters. In addition to using the OES system for its own measurements, WiPAL offers an opportunity to extend these calibrations to regimes relevant for laboratory plasma astrophysics.

\section{Major Experiments in WiPAL}

\subsection{Large Multi-Cusp Plasma Confinement}\label{sec:confinement}

WiPAL is the largest axisymmetric magnetic ring cusp ever constructed for confining plasmas. The permanent magnets effectively limit the loss area to $\sim1\%$ of the total vacuum vessel surface area, which
in turn leads to high, steady-state confinement. The combination of WiPAL's cusp confinement and diagnostic capabilities presents an opportunity for studying and characterizing the confinement of magnetic cusps.
The cusp field is created by 36 rings of alternating-polarity magnets, as shown in figure~\ref{fig:Cusp}. This field is localized to the edge, dropping off to the background earth field within 20 cm of the vessel wall. Plasma losses in this configuration are limited to a small cusp width on the face of each magnet. To electrically 
isolate the plasma from the vessel wall, each magnet is covered by a thin insulating ceramic tile. 

The long established empirical loss width for a magnetic cusp is the hybrid gyroradius $w\sim\sqrt{\rho_{i}\rho_{e}}$ \citep{Cusp}, yet newer studies have found a scaling with neutral pressure which suggests more detailed physics than was previously employed \citep{hubble2014}. Since WiPAL is able to create and sustain plasmas at very low pressures, this neutral pressure scaling can be investigated over several orders of magnitude. Accurate measurements of the ion velocity distribution at the cusp can be made with the Fabry-Perot system. Additionally, owing to the insulating boundary condition, {\it in situ} probes at the ceramic tile can directly measure the loss width. Understanding the loss width scalings in a magnetic cusp is important not only for predicting plasma confinement in WiPAL but also to a number of other applications such as plasma processing \citep{malik1994}, Hall-thrusters \citep{sengupta2009}, and neutral beam injectors \citep{stirling1979}. 

\subsection{Self-Exciting Dynamos}\label{sec:dynamo}

Astrophysical plasmas are often characterized by high magnetic Reynolds numbers ($Rm = V L/ \eta$, where $\eta$ is the magnetic diffusivity, $V$ is a characteristic flow speed, and $L$ is the system size) wherein turbulent, flow-dominated plasmas continuously transform kinetic flow energy into magnetic energy. Understanding this energy transformation and predicting what form the magnetic field might take, be it small-scale turbulent magnetic fields or large-scale magnetic flux, is the dynamo problem. The theory of dynamos has shown that the scales at which the magnetic energy grows are largely determined by the relative values of the magnetic Reynolds number and the fluid Reynolds number ($Re=VL/\nu$, where $\nu$ is the dynamic viscosity). The ratio $Pm=Rm/Re$, called the magnetic Prandtl number, is thought to be a critical parameter that governs the nature of many astrophysical phenomena since it sets the relative collisional dissipation scales of the fields and flows. Astrophysical plasmas span a wide range of $Pm$; diffuse plasmas have $Pm\gg1$ whereas denser plasmas have $Pm\ll1$. 

Dynamos can be classified as small-scale or large-scale. Small-scale dynamos tend to generate magnetic energy but little net magnetic flux, whereas large-scale dynamos generate both net flux and energy. While the process by which turbulence generates magnetic energy at small scales seems theoretically well understood \citep{schekochihin04,iskakov2007_prl}, understanding how a large-scale magnetic field self-organizes from small-scale magnetic fluctuations in astrophysical systems  remains a grand challenge for plasma astrophysics. 

\begin{figure}
  \centerline{
  \includegraphics[scale=1]{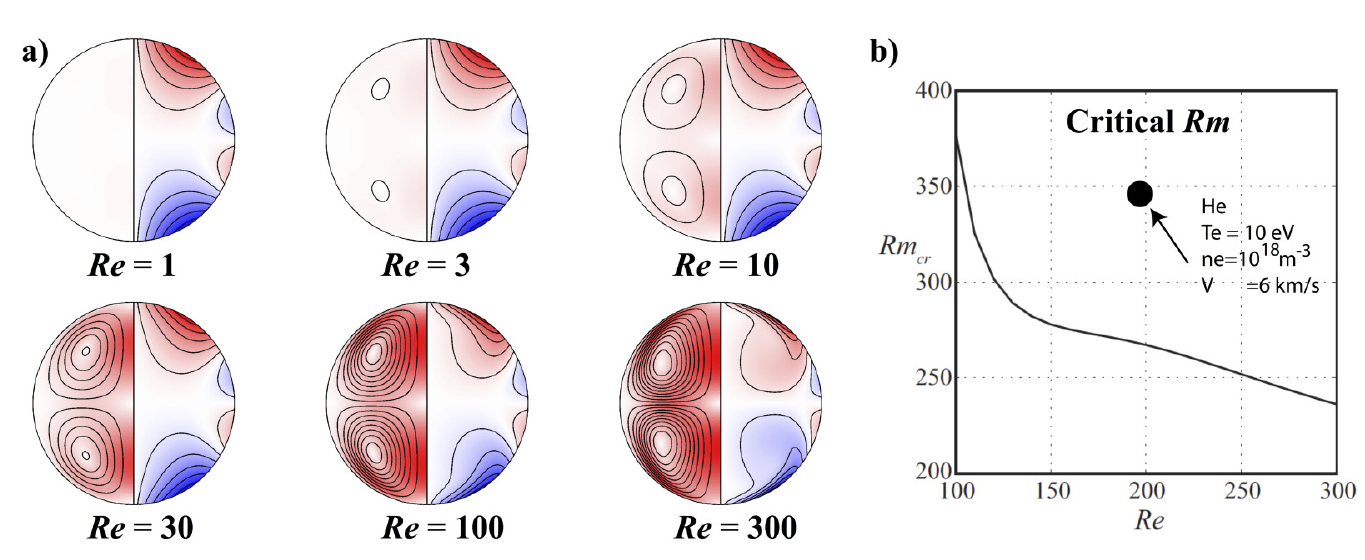}}
  \caption{a) Hydrodynamic simulations of edge-driven, two-vortex axisymmetric flow in He for increasing $Re$. Axis of rotation and symmetry is vertical; left hemispheres show poloidal flow stream lines and poloidal flow magnitude, right hemispheres show toroidal flow contours. Poloidal flow scale is amplified 4x relative to toroidal flow scale. b) Kinematic dynamo growth rate of the flows shown in (a). Increasing $Re$ results in a larger amount of kinetic energy in poloidal flows, which in turn lowers the critical $Rm$ required for positive magnetic eigenmode growth. Flow is only axisymmetric for $Re < 300$; hydrodynamic instabilities produce non-axisymmetric modes at higher $Re$, requiring full 3-D solutions to the induction equation.}
  \label{fig:MPDX}
\end{figure}

Studying plasma dynamos in the laboratory requires a previously
unexplored regime of laboratory plasmas. Unmagnetized, fast-flowing, and highly conducting plasmas are required so that magnetic fields can be stretched and amplified by the plasma inertia. Cast in dimensionless terms, dynamo action requires plasmas with high $Rm$ and high Alfv\'{e}n Mach number ($Ma=V/V_{A}\gg1$). The particular geometries being pursued in the MPDX configuration of
WiPAL build upon similar geometries used by liquid-metal experiments.
Mechanically stirred liquid-sodium experiments have observed spontaneous magnetic field generation \citep{monchaux2007,monchaux2009,gallet2012} and have added to our understanding of
astrophysical and geophysical dynamos \citep{Lathrop2011}. A plasma
experiment has the potential to extend these studies to parameters
more relevant to astrophysics. Beyond the obvious fact that most
naturally occurring dynamos are plasmas, the use of plasma rather than
liquid metals corresponds to magnetic Reynolds numbers increased by a
factor of 10 or larger. Additionally, viscosity can be varied
independently of the conductivity, with $Pm$ ranging from $0.1$ to $10$, reflecting the wide range found in different astrophysical systems.

Extensive modeling and theoretical work have been carried out to find flow schemes for MPDX
that excite a large-scale dynamo \citep{Spence.APJ.2009, Khalzov2013, Khalzov2012a, Khalzov2012,
Katz2012RSI}.  By biasing the cathodes in the magnetized edge,
torque is injected at the boundary of the plasma and is viscously coupled
to the core. Dynamo-relevant flows are found by solving the
Navier-Stokes equation for a particular boundary condition imposed by the
cathode drive. The resulting flow is then used to solve the kinematic
dynamo equation. One such flow that results in dynamo action is shown
in figure~\ref{fig:MPDX}. The imposed boundary condition drives two
counter-rotating vortices in each hemisphere with a small flow
direction reversal near the equator. For fairly modest $Rm$ and $Re$, positive dynamo growth is expected. Work on MPDX is currently directed at optimizing the flow drive and attempting to create this two-vortex flow. 

 \begin{figure}
  \centerline{\includegraphics[scale=1]{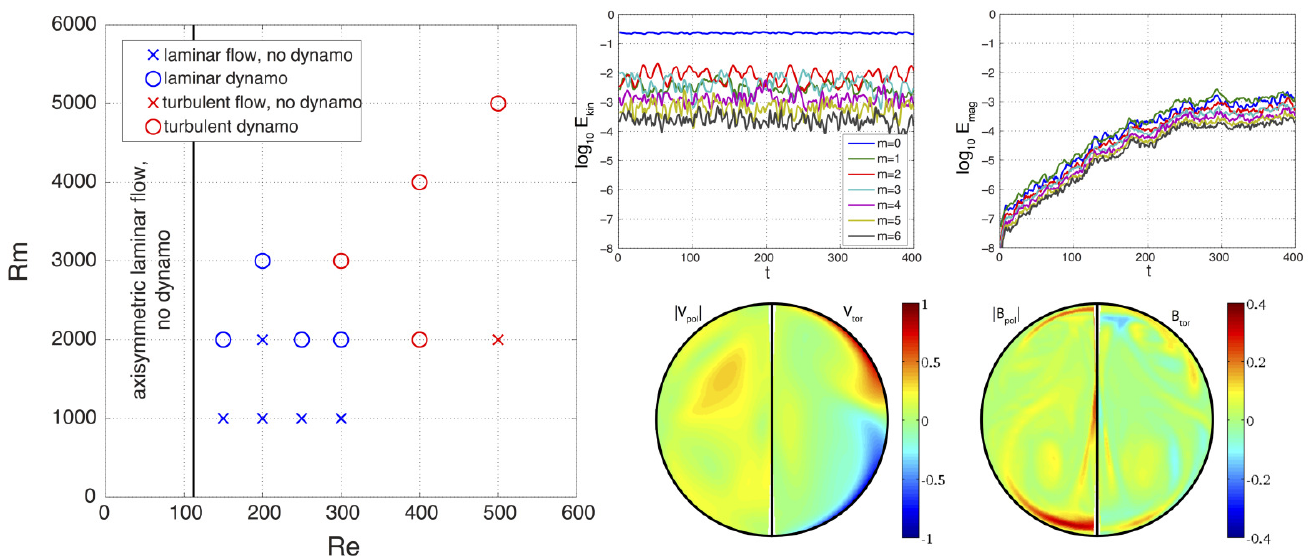}}
\caption{Numerical solutions for turbulence driven small scale dynamo and evidence for large scale field generation. 
To achieve $Rm=2000$ and $Re=400$ corresponds to a helium plasma with  $V_{max}$=10 km s$^{-1}$, $T_e$=30 eV, $T_i$=1.2 eV, $n_e=$ 10$^{18}$ m$^{-3}$, and $Z_{\rm eff}=$1.2. The left plot shows where simulations find turbulent dynamos in $Re$ and $Rm$ space.   }
\label{fig:Re_Rm_small_scale}
\end{figure}

Fast small-scale dynamos can be excited by chaotic flows with the high values of $Rm$ and $Pm$ possible in WiPAL. Chaotic flows with positive Lyapunov exponents could be achieved with large $Re$ turbulence or by using time-dependent and highly viscous laminar flows \citep{Khalzov2013}. Three-dimensional numerical simulations using turbulent flows where $Re\sim500$ and $Rm$ is very high have confirmed that turbulent, small-scale magnetic fields naturally develop in high-$Pm$ plasmas. Moreover, the magnetic amplification of these flows is roughly independent of $Rm$ as $Rm$ continues to increase, a hallmark of fast dynamos. Figure~\ref{fig:Re_Rm_small_scale} shows several example of flows where fast dynamo growth is possible. Simulations predict fast dynamo growth at plasma parameters that can be achieved with increased input power from ECR heating. Because $Rm$ is very large in most astrophysical systems, virtually any astrophysical dynamo must be fast. Understanding the conditions for fast dynamos and exploring whether they can generate magnetic fields on large scales is an open problem in plasma astrophysics. 

\subsection{Magnetic Reconnection}
\label{sec:TREX}

\begin{figure}
  \centerline{
  \includegraphics[scale=1]{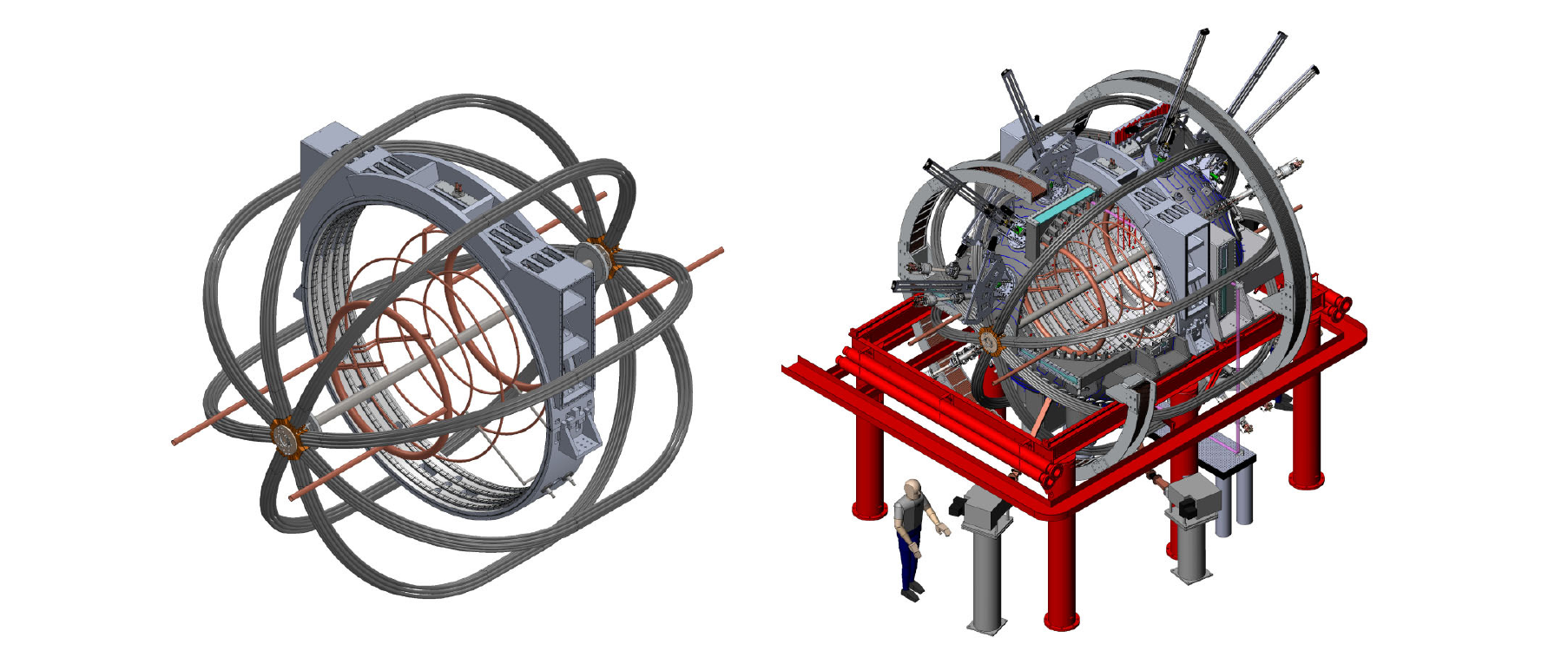}}
  \caption{Toroidal field coils and a cylindrical insert holding additional coils for driving reconnection (left) will be inserted into the WiPAL vessel. The large set of external Helmholtz coils seen in the figure (right) is a recent upgrade to the WiPAL facility with utility for many future experiments.}
  \label{fig:TREX1}
\end{figure}

\begin{figure}
  \centerline{\includegraphics[scale=1]{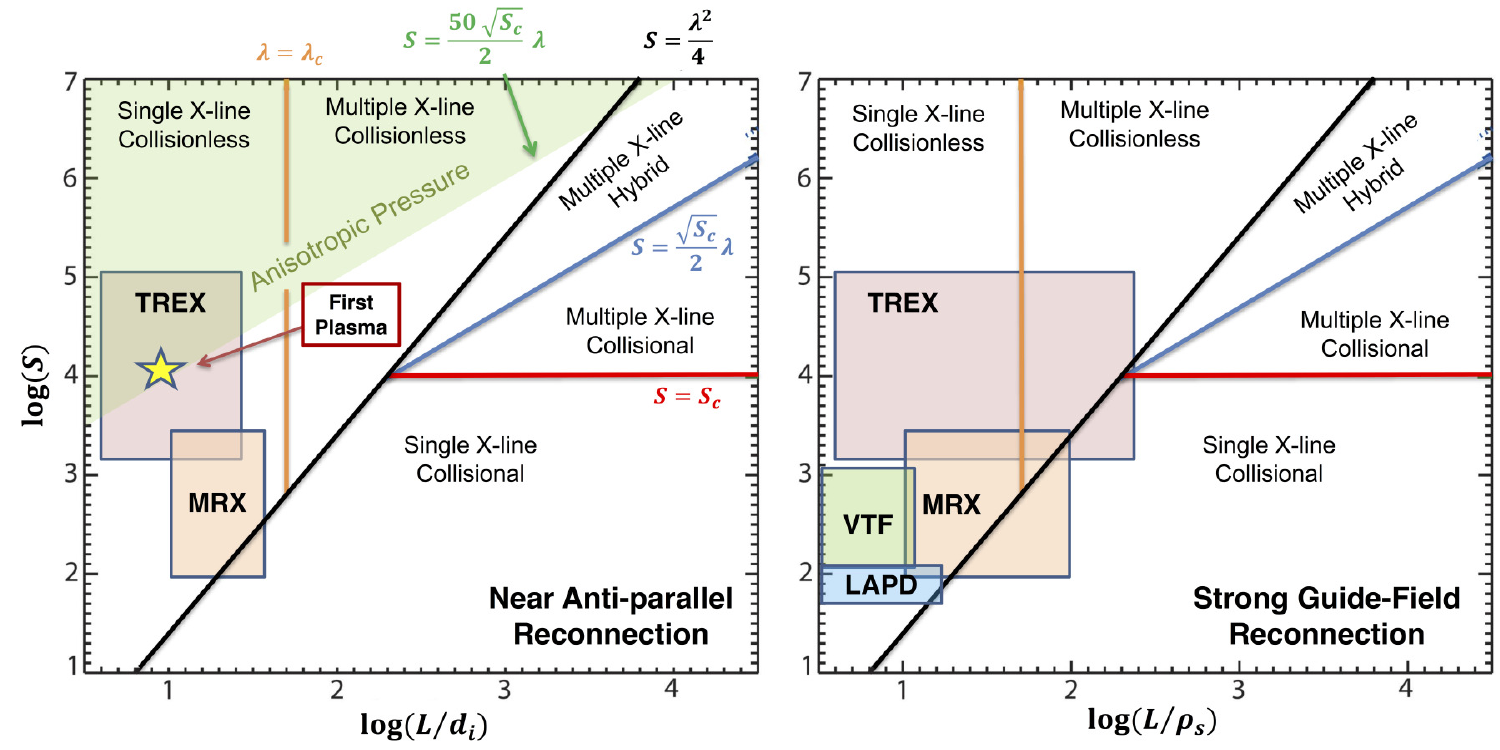}\hspace{0.2in}}
  \caption{Reconnection phase space diagrams for weak guide-field reconnection (left) and strong guide-field reconnection (right) identifying various regimes of reconnection. The star characterizes the first plasma obtained with TREX, demonstrating its ability to access the new collisionless reconnection regime.}
  \label{fig:TREX2}
\end{figure}

One of the fundamental processes in nearly all magnetized plasmas is magnetic reconnection: a change in the magnetic topology of the plasma that converts stored magnetic energy into particle kinetic energy \citep{priest2000}. Given its importance, magnetic reconnection has been studied extensively through theory, numerical computation, spacecraft observations, and laboratory experiments. In recent years, new frontiers have emerged with an emphasis on particle energization, reconnection with multiple X-lines in large systems, the role of kinetic effects in collisionless reconnection, and extending the evolution of reconnection from two dimensions to three. For experiments to remain relevant and to advance this maturing field, new devices are needed that access these regimes. To address this need, we are now operating the Terrestrial Reconnection Experiment (TREX), which is the largest dedicated reconnection experiment to date. In addition to WiPAL's existing vacuum vessel, plasma confinement, heating, and diagnostic suite, TREX consists of a cylindrical insert holding coils used for driving reconnection (shown in figure~\ref{fig:TREX1}). The parameter regimes expected for TREX are shown in table~\ref{tab:TREX}. The full operation of TREX within WiPAL renders the facility unique in its ability to address the expanding frontiers of reconnection research. 

An important component to the implementation of TREX into the WiPAL user facility is the cylindrical insert housing the reconnection drive coils and the addition of a toroidal field coil. The insert and TF coil allow TREX to access multiple magnetic configurations (e.g. antiparallel reconnection, strong guide-field reconnection, and 3D reconnection), as well as a wider range of reconnection regimes. TREX also utilizes a suite of probe arrays specially developed to characterize reconnection dynamics. These include stationary magnetic and electric probes along with sweep probe arrays that allow for a 2D region to be mapped out over repeated shots (\S \ref{sec:Probes}). The necessary dynamics can be captured on time scales up to 20 MHz with the diagnostic suite. For a desired value of the reconnecting field, the configuration must be driven at the corresponding loop voltage $V_{loop}=2{\pi}R(0.1v_{A}B_{r})$. For example, a loop voltage of 5 kV for a 150 G reconnecting field is required to access fast reconnection in a hydrogen plasma. This consideration has necessitated the development of reliable, high-power pulsing circuits utilizing class D ignitron switches to drive reconnection. These drive circuits can be pulsed as fast as every 10 seconds. 

So far, the term ``collisionless reconnection" has referred to systems where the electron and ion distributions can remain near Maxwellian, but the collisionality is sufficiently low that their collective fluid behaviors decouple at the ion scale and Hall currents become important. However, in a truly collisionless plasma, pressure anisotropy develops which strongly impacts the properties of the reconnection process in ways not accounted for in traditional Hall reconnection. In fact, spacecraft observations \citep{hwang2013} and kinetic simulations at the full ion-to-electron mass ratio \citep{le2015} show that large-scale current layers are driven by electron pressure anisotropy that builds in the reconnection region due to kinetic electron trapping effects. To maintain pressure anisotropy, the time between electron collisions must be long compared with the full transit time of a fluid element through the reconnection layer \citep{egedal2013,le2015}. 

\begin{table}
  \begin{center}
\def~{\hphantom{0}}
 \begin{tabular}{l || c | c | c | c | c |}
{}&$n_e (10^{18}\text{ m}^{-3})$&$T_{e}$ (eV)& $B_{r}$ (T)& $B_{g}$ (T) & L (m)\\
\hline
Terrestrial Reconnection Experiment&$0.1-10$&$8-40$&$0.04$&$0-0.3$&$0.8-1.8$\\
 (TREX; UW Madison)&&&&&\\
Magnetic Reconnection Experiment&$2-100$&$5-10$&0.03&$0-0.1$&0.3-0.8\\
 (MRX; Princeton)&&&&&\\
Versatile Toroidal Facility&$0.1-1$&$8-30$&0.01&0.1&0.3\\
 (VTF; MIT)&&&&&
\end{tabular}
  \caption{Key parameters for various reconnection experiments in
    hydrogen plasmas where $B_r$ and $B_g$ are the reconnecting and guide fields, respectively.}
  \label{tab:TREX}
  \end{center}
\end{table}

As a valuable tool for displaying the various regimes of reconnection and their transitions, \citet{daughton2012} developed the reconnection phase diagram spanned by the Lundquist number $S$ and the normalized system size $\lambda$ with respect to the ion sound Larmor radius, $\rho_{s}=\sqrt{m_{i}(T_{e}+T_{i})}/eB$, or the ion skin depth, $d_{i}=c/\omega_{pi}$, for strong and weak guide-field reconnection, respectively \citep{ji2011}. A convenient way of representing the constraint for anisotropic pressure on a system is the condition $S>10(m_{i}/m_{e})(L/d_{i})$. The anisotropic pressure region of this phase space is shown in figure~\ref{fig:TREX2}. As indicated by the star, TREX has already demonstrated its ability to access this regime of collisionless reconnection. TREX is able to experimentally study the role that electron pressure anisotropy has on particle heating. In addition, the narrow current layers driven by the anisotropy may in 3D be unstable to reconnection at oblique angles. This effect of 3D reconnection may be important to the self-consistent evolution and generation of reconnection with multiple X-lines. 

\subsection{Acoustic Waves, Helioseismology, and Angular Momentum Transport}
\begin{figure}
\centerline{\includegraphics[scale=1]{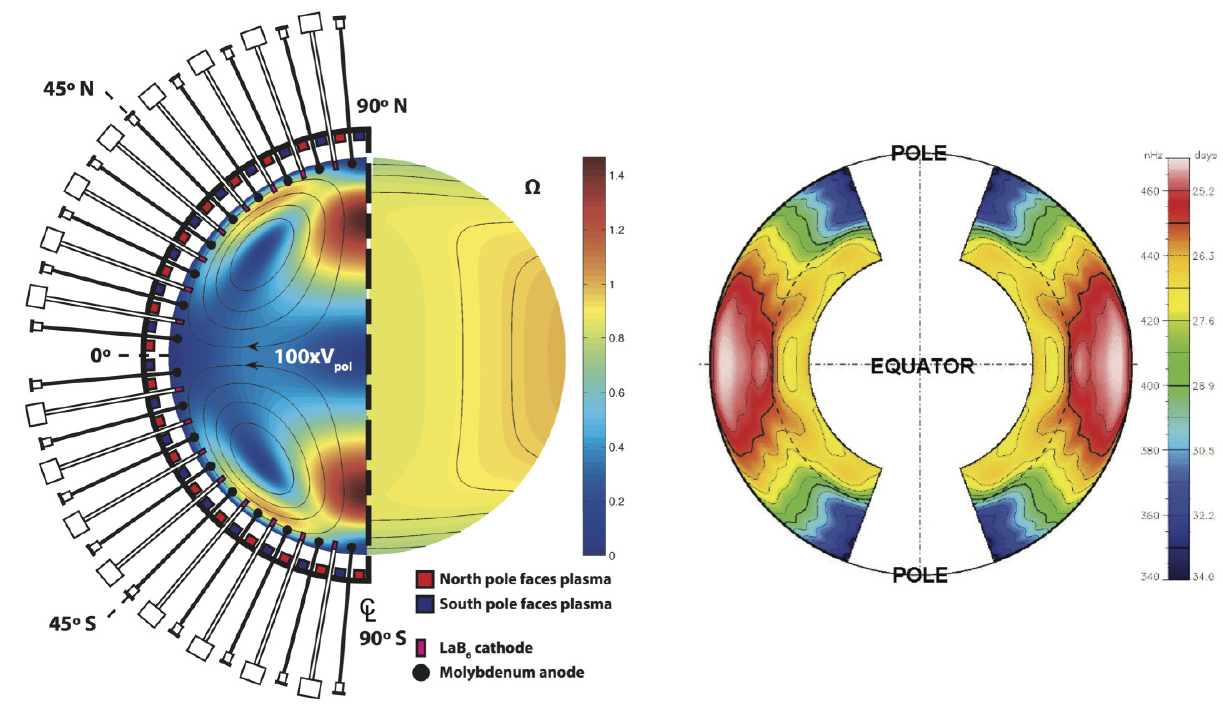}}
\caption{A simulation of solar-like flow inside the WiPAL vessel with streamlines of poloidal flow (left). The internal rotation profile of the Sun inferred from helioseismic inversions \citep{solarpic}.  Pink/red denote fast rotation and blue/green denote slower rotation as indicated by the color bar.  Inversions near the rotation axis are unreliable and are thus omitted from the plot.  The image of the angular velocity profile is reflected about the rotation axis in order to illustrate the full spherical geometry.}
\label{fig:solar}
\end{figure}

The WiPAL facility can study both basic solar plasma phenomena as well as longstanding mysteries explored by the heliophysics community, such as angular momentum transport in the Sun and other late-type stars.  It is well established from observations and modeling that late-type stars with convective envelopes spin down as they age due to torques exerted by magnetized stellar winds \citep{Mestel1968,barne07,matt15}.  It is less well understood how this angular momentum loss is transmitted through their convection zones and into their radiative cores.  Helioseismic probing of our own Sun suggests that this coupling is efficient in the sense that the radiative core currently has a rotation rate that is comparable to the convective envelope \citep{thomp03}. However, these same helioseismic inversions reveal that the radiative core is rotating nearly uniformly while the convection zone rotates differentially, with a $\sim 30$\% decrease in angular velocity from equator to pole \citep{thomp03}.  The transition between these two distinct rotation regimes is called the solar tachocline, which overlaps with the base of the convection zone and the convective overshoot region.

Angular momentum transport within the solar convection zone is attributed to turbulent magnetized convection and occurs on convective time scales of weeks to months.  By contrast, the transport of angular momentum across the solar tachocline, coupling the core and envelope, is thought to occur on much longer time scales of millions to billions of years \citep{spieg92}. Our current understanding of both short-term and long-term angular momentum transport in the Sun relies on incomplete and indirect observational data (e.g. helioseismic inversions) and numerical magnetohydrodynamic (MHD) models. WiPAL offers an opportunity to study some aspects of the relevant dynamics in the laboratory for the first time.

Using the drive system as MPDX, a solar-like flow can be programmed at the boundary of the WiPAL vessel. A calculation of the structure of this flow is shown on the left of figure \ref{fig:solar}. This flow is aimed at mimicking the helioseismic inversion, shown on the right of figure \ref{fig:solar}, with strong toroidal rotation near the equator. Careful scrutiny of how such a boundary-forced differential rotation spreads into the interior through meridional flows, viscous diffusion, thermal gradients, and magnetic tension will provide insight into core-envelope coupling and the resulting spin-down of late-type stars.  Magnetic tension in particular is thought to be responsible for maintaining the uniform rotation of the radiative interior by suppressing shear, as described by  Ferraro's theorem \citep{macgr99}. Furthermore, inserting a dipole electromagnet into the core of the WiPAL vessel will create a mock tachocline, testing the structure and stability of the boundary layer proposed by magnetic tachocline confinement models \citep[e.g.,][]{gough98}.

By exciting the acoustic normal mode spectrum of a spherical plasma with external antennas, the stationary plasma response can be characterized. It is predicted by theory and observed in the Sun that these normal modes split when rotation is introduced \citep{CD2002}. Helioseismology relies on these mode splittings at different locations on the Sun to discern the flow profiles of the interior. A similar analysis is available at WiPAL with the addition of the external magnetic probe array described in \S\ref{sec:Probes}. When differential rotation is imposed at the boundary, the resulting spectra can be spatially decomposed and compared to the stationary mode structure. A mathematical inversion can then be used to propose likely candidates for internal global flow profiles, which can be benchmarked by Mach probe measurements under similar conditions. This would form the basis of a minimally invasive, global flow profile diagnostic at WiPAL akin to helioseismic inversions.

Using this helioseismic diagnostic and solar flow drive, WiPAL will study the short-term angular momentum transport in the solar convection zone. Experimental set-ups, such as that shown in figure \ref{fig:solar}, can be used to investigate the physics of turbulent transport. In the Sun, turbulent angular momentum transport is thought to be responsible not only for sustaining the solar differential rotation, but also for regulating the amplitude and structure of the meridional flow by means of gyroscopic pumping \citep{miesc11}.  More generally, turbulent angular momentum transport, mean flows, and thermal gradients in the solar convection zone are all thought to be intimately linked through nonlinear feedbacks that can be explored with WiPAL. 

\subsection{Centrifugally Driven Stellar Winds}

Plasma wind launched from star surfaces carries stellar magnetic field into the local heliosphere and out to the interstellar medium. This system is complex and is governed by a range of processes, including magnetic reconnection, turbulence and particle heating, all topics of present-day heliospheric research. One aspect that has received little experimental attention is the magnetic topology of the advected magnetic field in the interface region between the magnetically dominated corona and the flow-dominated wind. This region is critical both in creating the Parker spiral and in determining the mass loss rates of stars. The Centrifugal Wind Experiment (CWE) at WiPAL will explore centrifugal breakout of wind from spinning dipolar magnetospheres. 

The winds of rapidly rotating giant stars and pulsars are particularly relevant. For rapidly rotating giant stars, centrifugal breakout has been predicted to proceed in episodic bursts of plasmoids \citep{ud-Doula06}. Recent observations by the Microvariability and Oscillations of STars (MOST) telescope have called into question the validity of this breakout model, suggesting instead a smooth outflow \citep{Townsend2013}. Modeling centrifugal breakout in a laboratory setting can advance our understanding of this phenomenon. Pulsars have been extensively modeled using analytical MHD and 3D particle-in-cell (PIC) simulations of both aligned and obliquely rotating magnetospheres \citep{Spitkovsky06,Philippov2014}. Both cases predict the creation of a magnetic Y-point where the closed magnetic topology breaks into an open configuration. This is expected to occur at the Alfv\'{e}n radius, the location where the kinetic and magnetic energy densities are equal. WiPAL'€™s large volume, well-developed power supplies, and comprehensive diagnostics make it suitable for experimental analysis of these active areas of research.

\begin{figure}
	\centerline{\includegraphics[scale=1]{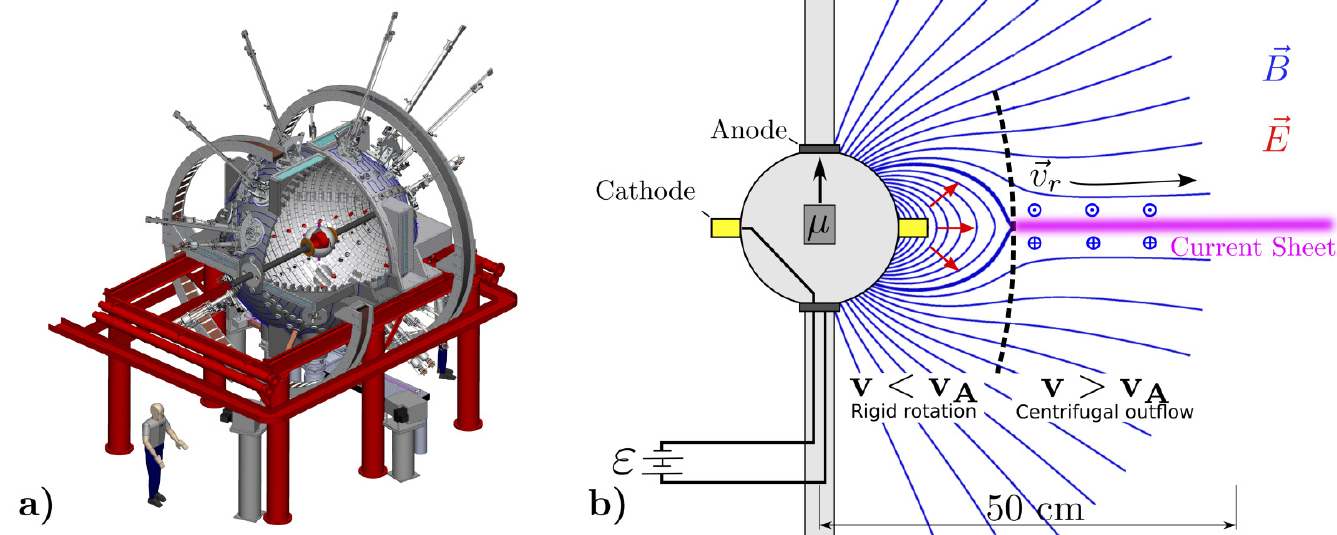}}
	\caption{(a) WiPAL facility with centrifugal wind experiment installed and (b) proposed configuration showing the electromagnetic stirring scheme and Y-point.}
\label{fig:Wind}
\end{figure}

A laboratory analog of a rotating, stellar magnetosphere is possible using the electrostatic stirring techniques developed for MPDX with the addition of a dipole magnetic field at the center of the vessel. Through ECR heating, the dipole magnetosphere will be filled with confined plasma. Biased electrodes at the equatorial surface of the dipole will establish a cross-field potential gradient, leading to ${\bf E}\times{\bf B}$ stirring. As the plasma spins to critical velocity, a centrifugal wind will be launched. In this way, we can use a large cross-field potential to spin the magnetosphere and model the above astrophysical situations. 

In preliminary experiments, an electrically insulated, spherical SmCo magnet ($r\simeq10$ cm, surface field $\sim4$ kG) will serve as the dipole source, shown in figure \ref{fig:Wind}. ECR heating will produce plasma on the spherically symmetric 875 G surface. Using helium, expected temperatures and densities will be similar to prior MPDX experiments ($T_{e}\approx5-20 eV$ and $n_{e}\approx10^{12}$ cm$^{-ˆ'3}$). Equatorial cathodes, spaced $\approx10$ cm apart, will be biased up to 1 kV to establish a cross-field potential, producing axially aligned ${\bf E}\times{\bf B}$ rotation. 

WiPAL's existing diagnostics can be used for investigation of the Y-Point and current sheet regions. The standard suite of Langmuir and Mach probes will be used for basic plasma measurements at the edge and will supplement the non-invasive measurements. A 2D, motor-controlled, 20 MHz resolution magnetic probe array will serve as the primary diagnostic. The magnetic probe will look for the characteristic Y-point opening of the magnetic field lines at the Alfv\'{e}n radius and will explore the equatorial current sheet at larger radii. 

Planned upgrades will explore other regimes. For example, obliquely rotating pulsars generate larger spin-down power than aligned rotators \citep{Spitkovsky06} and provide a more realistic setting for studying current sheet reconnection. By supplementing the dipole magnet with phased coils, an obliquely rotating magnetosphere can be created. This will allow for experimental confirmation of the above prediction and characterization of the change in the undulating current sheets.

\subsection{Stability of Astrophysical Jets in Background Plasma}
\label{sec:jets}

Astrophysical jets are collimated magnetized plasma outflows from accreting bodies such as active galactic nuclei, binary systems, and young stellar objects. It is thought that jets could play a key role in angular momentum transport in accretion systems. Jets are launched and collimated in regions much smaller than the current resolution limits of observations, so the details of this process remain elusive. Certain models of the launching process predict that astrophysical jets arise due to magnetic fields in the accretion disk \citep{Lovelace1976,1982MNRAS.199..883B,Spruit2010}. In these models, a dipole-like magnetic field is sheared by Keplerian rotation in the disk. This creates a large electric field between the center and edge of the disk, driving current along the dipole field. Then ${\bf J}\times{\bf B}$ forces accelerate and collimate plasma, creating the jet structure. These jets should be susceptible to current-driven instabilities, but recently published theories suggest that the presence of external plasma pressure promotes collimation and enhances the stability of the jet structure out to large distances \citep{2003MNRAS.341.1360L, Li2001,Li2006}.

\begin{figure}
   \centerline{ \includegraphics[scale=1]{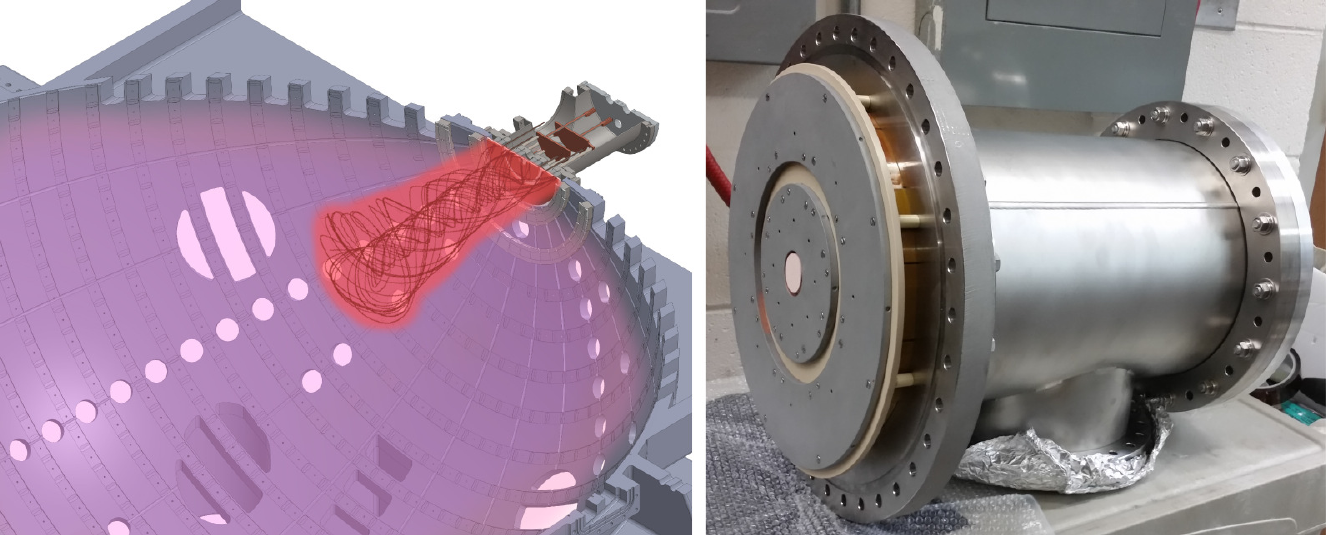}}
  \caption{Experiment to study Jet expansion into background, finite pressure plasmas. Simulation of Jet is taken 
  from \cite{Li2006} and superimposed into WiPAL plasma schematic, showing coaxial injection geometry.
  Right:  photo of the assembled pulsed jet source  preparing for first tests.}
  \label{fig:PMPJS}
\end{figure}

The Permanent Magnet Pulsed Jet Source (PM-PJS), shown in figure~\ref{fig:PMPJS}, creates an astrophysical jet-like plasma utilizing a design similar to that used by Hsu and Bellan \citep{Hsu2002}. A large voltage potential is created between two coplanar annuli (approximately 30 cm in diameter). This simulates the potential created via the aforementioned accretion disk shearing action. The background magnetic field in the astrophysical system is simulated using two antisymmetric rings of permanent magnets. The use of permanent magnets instead of electromagnets allows the source to maintain WiPAL's cusp-confining boundary condition. Neutral gas is injected via internal channels in both annuli with ports placed at eight azimuthally symmetric pairs of magnetic field line foot points. This gas is ionized by the large voltage creating eight filamentary loop structures.  These loops merge and collimate, driving a magnetic jet into the center of WiPAL.

This experiment complements Caltech experiments \citep{Hsu2002, Hsu2003, Moser2012} by studying the evolution of a magnetic jet evolving into a vacuum.  Additionally, WiPAL allows the creation of a magnetic jet that propagates into a high-$\beta$ background plasma. This will help clarify the relationship between jet stability and external plasma conditions \citep{2003MNRAS.341.1360L, Li2001, Li2006}. Future work will focus on the shock and precursors that are formed between the magnetized jet and the unmagnetized background plasma as predicted in \cite{Li2006} and on how the development of the kink instability could depend on the strength of the background pressure. Experimentally it will be possible to vary the target plasma pressure by more than two orders of magnitude. The properties of the expanding magnetic plume (speed, density, magnetic field) as well as the shock precursors will be measured using scanning probes similar to those used in the reconnection and centrifugal wind experiments.

\subsection{Helicity Injection and Decaying Turbulence}
The generation of large-scale magnetic fields is a fundamental feature of many astrophysical systems. This dynamo action (as described in \S\ref{sec:dynamo}) is often attributed to a turbulent upscale transfer of small-scale magnetic helicity, $H=\int{\bf A}\cdot{\bf B}\,d\mathcal{V}$. The upscale transfer process is ideally described as a local, self-similar inverse cascade where helicity is conserved and transferred to larger scales \citep{frisch1975, Ji1999, Blackman2006}. In real systems however, the transfer process is much more complex and can involve non-local transfer directly from the small to large scales as described in the turbulent alpha effect. The upscale spectral transfer of magnetic helicity has been explored extensively via direct numerical simulations \citep[e.g.][]{alexakis2006}. Understanding this upscale transfer of magnetic helicity is key to explaining the creation of the large-scale magnetic fields observed throughout the Universe.

In addition to upscale spectral transfer of helicity, transport in space is necessary for large-scale dynamo action. Large-scale dynamos can be sustained in systems with boundary conditions allowing the outflow of helicity. Roughly speaking, ejection of one sign of magnetic helicity allows helicity of the opposite sign to grow without violating helicity conservation. This idea was originally suggested on theoretical grounds by \citet{blackman2001} and \citet{vishniac2001}. Later, it was shown that magnetic eruptions associated with solar activity transmit a net helicity flux, suggesting that the solar dynamo meets these boundary conditions \citep{rust2002,kusano2002, liu2014,pevtsov2014}. In addition to the solar dynamo, this boundary condition effect is thought to occur in galactic dynamos driven by supernovae \citep{rafikov2000_mnras}. There is evidence from numerical simulations that the generation of large-scale fields is promoted by boundary conditions that permit escape of magnetic helicity \citep{brandenburg2004, kapyla2010, hubbard2012}. Yet none of these ideas have been tested in the laboratory.

In the plasma physics community, helicity injection is primarily explored as a startup mechanism for magnetically dominated tokamak plasmas \citep{raman2003, raman2010, battaglia2009}. Some laboratory astrophysical experiments that inject net helicity into plasma systems have also focused on magnetic reconnection and other instabilities that occur during Taylor relaxation of low-$\beta$ plasmas \citep{cothran2009, gray2010, Jarboe2005}. All of these experiments rely on helicity injection at large scales (comparable to the scale of the plasma volume) relaxing via magnetic instability to minimum-energy, helicity-conserving Taylor states in low-$\beta$ plasmas \citep{Taylor1986}. As a complement to these studies, we will inject helicity at small scales and use the large size of the WiPAL vessel to directly observe the upscale spectral transfer and transport of helicity in a high-$\beta$, dynamo-relevant plasma conditions.

\begin{wrapfigure}{r}{2in}
\includegraphics[scale=1]{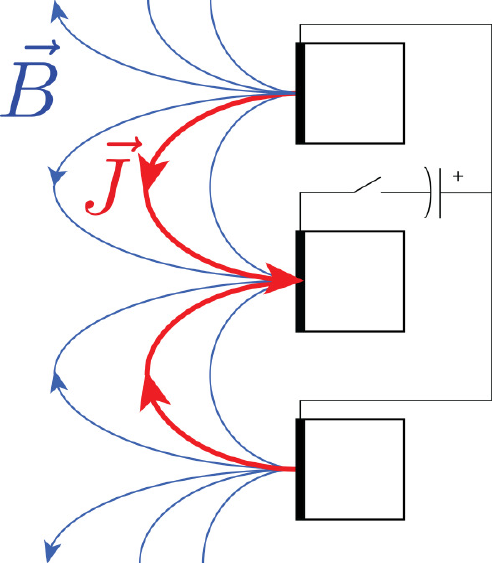}
	\caption{The axisymmetric ring cusp lends itself to a novel helicity injection experiment where currents, flowing between the magnetic pole faces can be configured to inject small scale magnetic helicity.}
	\vspace{-0.1in}
\label{fig:helicity}
\end{wrapfigure}

Generalizing the PM-PJS experiment discussed above (\S\ref{sec:jets}), helicity can be injected using the edge-localized cusp field inside WiPAL. A large potential can be applied between alternating rings, which will drive a current along cusp field lines and lead to an injection of helicity into the system without a guide field present (figure \ref{fig:helicity}) . By using all of the rings of permanent magnets inside the vessel, this helicity will be injected at a small ($\ell$=18, $m=0$) scale relative to the size of the vessel. The resulting rings of flux will then be blown off the wall into the bulk unmagnetized plasma with kinetic and magnetic energies near equipartition. In order to drive an upscale transfer instead of simple resistive diffusion, we require the resistive decay time to be much longer than the Alfv\'{e}n time of this system. Cast in dimensionless terms, this means that the Lundquist number, $S\equiv \tau_{r}/\tau_{A}$, must be large. Using the high-power capacitor banks created for TREX, magnetic fields of $B\simeq100$ G can be induced from current driven along the cusp field lines. For WiPAL discharge parameters, this corresponds to $S\simeq1000$. Under these conditions, we expect helicity to undergo a turbulent transfer before resistive diffusion can dissipate the injected magnetic energy. Observing the transfer and transport of helicity in this set-up will complement dynamo studies conducted by MPDX and provide the astrophysics community with a laboratory environment to probe this fundamental process.

\section{Conclusion} 

The Wisconsin Plasma Astrophysics Laboratory (WiPAL) provides the plasma astrophysics and fundamental plasma physics communities with a unique opportunity to study plasma phenomena in a laboratory setting. Hot, dense, unmagnetized, and fully ionized plasmas are routinely created and confined in quiescent states for seconds as astrophysics experiments are performed. This user facility has been designed and operated with the goal of maximizing both plasma performance and flexibility of use. 

To date, WiPAL has confined steady-state plasmas which are hotter and denser than any other large-scale non-magnetically dominated plasma device. These parameters have proven to be sufficient for both studying dynamo relevant regimes (MPDX) and providing high-Lundquist-number target plasmas for reconnection studies (TREX).  In the longer term, the heating power will more than double as the cathode system is completed and the ECR system is installed.

Diagnosing the high-performance unmagnetized plasmas in WiPAL has been an area of intense focus. Arrays of robotically controlled magnetic and electrostatic probes are capable of mapping out large areas of WiPAL plasmas. Advanced optical diagnostics have been developed with the goal of reducing the additional loss area  added by this suite of {\it in situ} probes. Millimeter-wave technology is used to power a compact heterodyne interferometer system for measuring electron density with high resolution. A Fabry-Perot interferometer is used to extract the ion velocity distribution. Finally, a complement of low-cost spectrometers coupled with collaborative modeling makes reliable estimates of the electron temperature and ionization fraction. All of these diagnostics provide a valuable set of data for every WiPAL discharge regardless of experiment. 

Perhaps most significantly, the WiPAL facility has already demonstrated the ability to quickly change experimental configurations.  This paper has outlined two major experiments already running on WiPAL (MPDX and TREX) as well as several new experiments in various stages of planning and implementation. We envision WiPAL transitioning in the near future to a user facility model in which investigators from outside the collaboration groups listed above could apply for and receive support to carry out experiments with technical support from the WiPAL staff.

The construction of the facility was supported by a National Science Foundation (NSF) Major Research Instrumentation grant. The MPDX and TREX research is now supported by NSF and the U.S. Department of Energy (DoE) and the NSF Center for Magnetic Self Organization in Laboratory and Astrophysical Plasmas (CMSO). The helioseismology studies are part of a collaboration with NCAR (M. Miesch) and are supported by a NASA graduate fellowship (E. Peterson). The stellar wind experiment is a collaboration with Princeton University (A. Spitkovsky) and is supported by an NSF graduate fellowship (D. Endrizzi).

\bibliographystyle{jpp}

\bibliography{WiPALJPP}

\end{document}